# Predictive Health Analysis in Industry 5.0: A Scientometric and Systematic Review of Motion Capture in Construction

Md Hadisur Rahman[1*], Md Rabiul Hasan[1], Nahian Ismail Chowdhury[1], Md Asif Bin Syed[1], Mst Ummul Farah[2]

1 Department of Industrial and Management Systems Engineering, West Virginia University, Morgantown, WV 26506, USA

2 Bangladesh University of Textiles, Dhaka 1208, Bangladesh

mr00063@mix.wvu.edu, mh00071@mix.wvu.edu, nahiyan313@gmail.com, ms00110@mix.wvu.edu, ummulfarah894@gmail.com

*Correspondence: mr00063@mix.wvu.edu; Tel.: +13042829952

## Abstract

In an era of rapid technological advancement, the rise of Industry 4.0 has prompted industries to pursue innovative improvements in their processes. As we advance towards Industry 5.0, which focuses more on collaboration between humans and intelligent systems, there is a growing requirement for better sensing technologies for healthcare and safety purposes. Consequently, Motion Capture (MoCap) systems have emerged as critical enablers in this technological evolution by providing unmatched precision and versatility in various workplaces, including construction. As the construction workplace requires physically demanding tasks, leading to work-related musculoskeletal disorders (WMSDs) and health issues, the study explores the increasing relevance of MoCap systems within the concept of Industry 4.0 and 5.0. Despite the growing significance, there needs to be more comprehensive research, a scientometric review that quantitatively assesses the role of MoCap systems in construction. Our study combines bibliometric, scientometric, and systematic review approaches to address this gap, analyzing articles sourced from the Scopus database. A total of 52 papers were carefully selected from a pool of 962 papers for a quantitative study using a scientometric approach and a qualitative, in-depth examination. Results showed that MoCap systems are employed to improve worker health and safety and reduce occupational hazards. The in-depth study also finds the most tested construction tasks are masonry, lifting, training, and climbing, with a clear preference for marker-less systems. One significant finding highlights the types of MoCap technology that focus on "technology and design improvement," then "safety and risk management." The studies examined feature a wide range of sample sizes, from 1 to 126 participants, with IMU sensors being the most used. Finally, the paper discusses future research directions, contributing significantly to the existing body of knowledge in applying MoCap within the construction workplace and aligning seamlessly with the principles of Industry 4.0 and 5.0.

## 1. Introduction

Motion Capture (MoCap) systems are advanced technological solutions designed to digitally record the patterns, gait, and human movement in three-dimensional space [1]. MoCap systems provide precise data on position, orientation, and a comprehensive phenomenon of movements employing various sensors, cameras, and techniques [2–4]. Several types of MoCap systems available in the market, each of which utilizes distinct technologies and operational methods. The optical MoCap system is classified into two categories: marker-based and marker-less. It relies on optical sensors such as cameras, which are equipped with specialized features (Infrared (IR) or visible light markers) to monitor the position and orientation of markers for marker-based systems [5–8]. Marker-less optical MoCap systems track and record motions without using physical markers attached to the human body, employing cameras with depth-sensing or image recognition capabilities [9]. On the other hand, the non-optical MoCap system employs other types of sensors, such as inertial sensors which combine accelerometers, gyroscopes, and magnetometers [10]. Magnetic and electro-magnetic MoCap systems operate using magnetic and electro-magnetic fields, respectively. These non-optical MoCap systems are applicable where the optical MoCap systems might be impractical due to some challenges and limitations, for instance, occlusions, restricted to the line-of-sight, indoor environment, or failure to operate in real-time MoCap. The MoCap systems, which include Inertial Measurement Units (IMUs), are lightweight, portable, can operate remotely without any wire connections, and may overcome those limitations and challenges [3]. Therefore, recently, researchers have focused on IMU-based MoCap systems to track and monitor human motion in different industrial workplaces such as sports, entertainment, gaming, construction, aerospace, automotive, robotics, healthcare, and safety [11–13]. Moreover, hybrid and versatile MoCap systems combining two or more MoCap technologies have emerged to overcome those challenges and improve data accuracy and precision [14–16].

The versatility of MoCap systems has led them to be employed in a wide range of applications [4,17–21]. Firstly, in medical and healthcare, MoCap systems help to diagnose and treat movement disorders [22], such as Parkinson's disease, cerebral palsy, and stroke can be diagnosed, and doctors can respond accordingly to the medications [23–26]. Other applications include rehabilitation programs, assessment of surgical outcomes, and reviewing a patient's motor functions, and protecting them from occupational injuries. Furthermore, previous research has investigated different approaches to protect patients from health hazards, including those caused by the COVID-19 pandemic, to ensure safety in healthcare environments [27]. In the automotive industry, MoCap systems can be applied to track the movements of workers in assembly lines, optimize the design of workstations, and develop new safety features [28–30]. MoCap applications in the aerospace industry include training astronauts, simulating the assembly of spacecraft, and developing new maintenance and repair services [1,31,32]. Manufacturing industries employ MoCap technologies to track the movement of robots and workers on production floors, optimize processess and improve quality [10,33]. Moreover,

MoCap systems are also used to maximize packing operations and enhance safety and efficiency in the logistics industry [34]. The entertainment and gaming industry also predominantly uses these technologies [35–40]. Finally, construction industries have benefited from this technological advancement while identifying ergonomic hazards, risks of falling and collision, and developing training programs and safety protocols [41]. Reducing ergonomic hazards and ensuring workers' health and safety are two of the most utilized applications of MoCap systems in construction workplaces [42–45].

With the evolution of Industry 4.0 and Industry 5.0 concepts, the application of MoCap technologies has become more convenient and cost-effective to ensure not only workers' health and safety but also increase the efficiency and productivity of worksites. Therefore, it is crucial to track technological advancement, progress, and trends of publications over the past few decades. A few reviews were conducted on the applications of MoCap systems other than construction fields [46–50]. One previous review paper focused on MoCap in general industrial applications, which includes the construction industry [1]. The study [1] only focused on an in-depth analysis of previous articles from 2015 to 2019. Another study [51] explores the crucial role of emerging technologies in addressing ergonomic hazards in construction workplaces during the Industry 4.0 era. The study focuses on the use of wearable sensors, extended reality technologies such as virtual reality (VR), augmented reality (AR), mixed reality (MR), and exoskeletons, and robotics as effective measures to minimize non-fatal injuries in construction workplaces. As the construction workplace is one of the biggest industries, requiring physically demanding tasks and increasing the risk of work-related musculoskeletal disorders (WMSDs), the application of MoCap technologies has emerged to overcome those challenges in this workplace.

However, there has yet to be a comprehensive review that investigates both scientometric and systematic reviews to explicitly examine the application of the MoCap systems in the construction workplace. The scientometric analysis is a quantitative and data-driven approach to examine scientific literature, citations, trends, patterns, and relationships within a particular field [51–55]. Thus, the bibliometric, scientometric, and in-depth analyses employing MoCap technologies in construction industries have not yet been systematically investigated. Therefore, a clear gap exists in the previous literature of understanding the publication trends, scientometric and in-depth analyses of MoCap systems applied in construction workplaces. To fill this gap, this systematic review paper employed bibliometric, scientometric, and in-depth analyses. The purpose of the paper is to report the publication trends, keywords co-occurrence analysis, co-authorship analysis, citations by document analysis and bibliometric coupling analysis. Following that, an in-depth review provides insight into MoCap technologies in construction. Finally, we aim to provide the limitations and future research directions based on the findings of this review paper. The broad scope of this study makes it appealing to a wide range of stakeholders, including those with a stake in MoCap from various industries, including research, education, construction, healthcare, manufacturing, and other relevant sectors. This study is structured as follows: the materials and methodology are described in Section 2, and the results, including scientometric and thorough evaluations of the literature, are presented in Section 3. The discussion is covered in Section 4, and Section 5 contains the future research directions. Lastly, Section 6 provides the conclusions of the review paper.

## 2. Materials and Methods

The study focuses on bibliometric, scientometric, and in-depth analyses to review articles pertaining to the application of MoCap systems in construction workplaces. Bibliometric (phase 1) and scientometric analyses (phase 2) were conducted using VOSviewer software (version 1.6.20). For in-depth analysis (phase 3), we carefully read the selected articles and pointed out the key findings in Microsoft Excel sheets. Following that, we then analyzed the insights and key features in the results and discussion sections (phase 4). The overall research procedures are depicted in Figure 1.

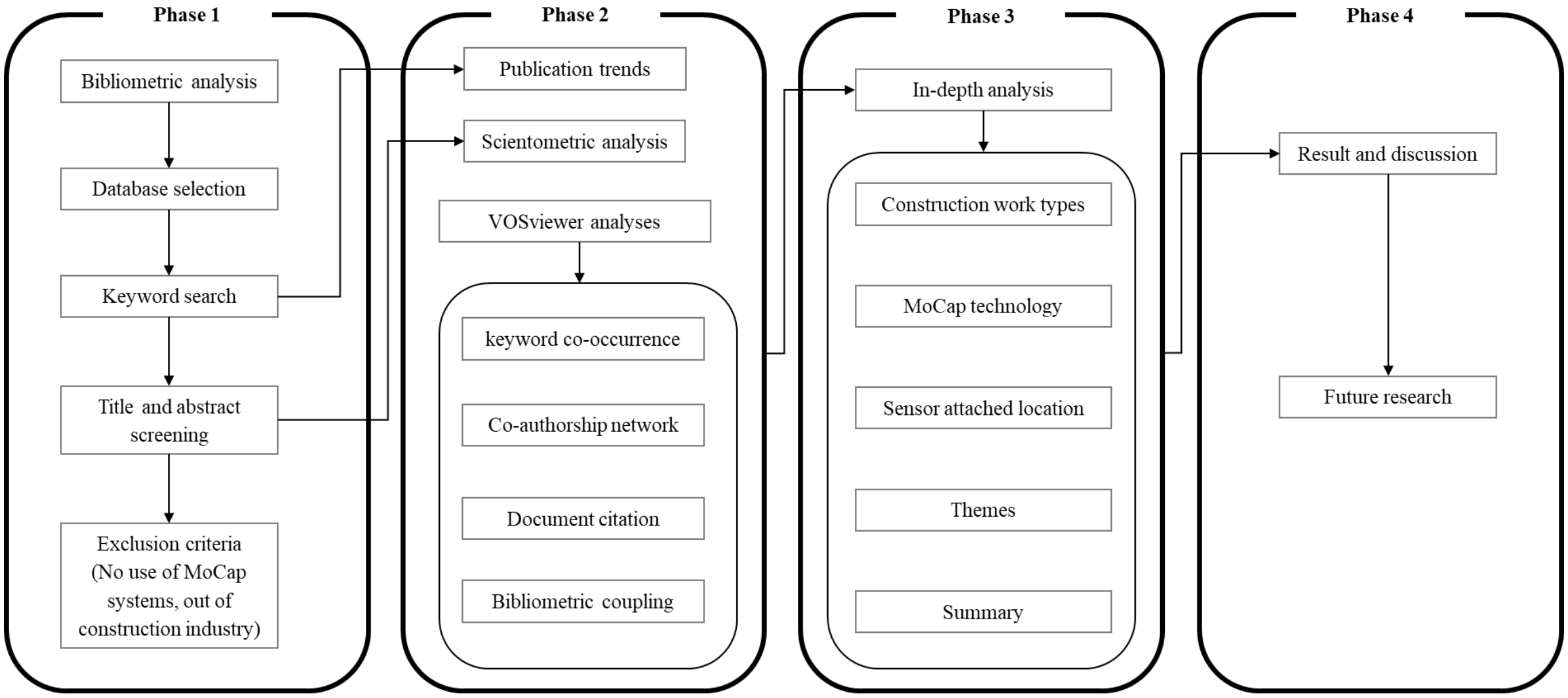

Figure. 1: Overall research methodology

### 2.1. Search Strategy and data extraction

We utilized the Scopus database as our primary source to collect a comprehensive compilation of literature on MoCap systems in the construction workplace. We chose Scopus for its extensive coverage and well-established reputation as a highly reliable and exhaustive bibliographic database [56,57]. Scopus is widely known for its comprehensive coverage of peer-reviewed journals, conference papers, and patents, providing an unmatched depth and breadth in its inclusion of scientific, technical, medical, and social sciences literature. We carefully designed our search strategy to encompass the broad range of MoCap technology used in the construction industry. We used a combination of specific phrases: *((MoCap*) OR (“motion data*”) OR (“motion* *track*") OR ("motion record*")) AND (construction*).* The search query was designed to cover the wide range of names and terms used in MoCap systems to include a complete selection of relevant research. Using this search approach in March 2023, we obtained a total of 962 documents. The metadata of these papers was extracted in CSV format, enabling a systematic and structured approach to data analysis. Extracting data simplifies the process for later Preferred Reporting Items for Systematic Reviews and Meta-Analyses (PRISMA) framework, scientometric analysis, and in-depth review.

The PRISMA framework is crucial for ensuring a meticulous and precise methodology in systematic reviews, which is essential for accurately synthesizing research findings [58–60]. We initiated our study using the PRISMA framework by conducting an extensive search in Scopus, as shown in Figure 2. This search yielded a total of 962 records. After carefully scrutinizing the titles and abstracts, we identified 113 relevant records. Following full-text evaluations, we narrowed the selection to 52 studies suitable for detailed analysis. To prevent any potential risk of biases in our review paper, we adopted a collaborative approach, where three authors reviewed the documents and categorized them based on their alignment with the scope of our research. The articles were classified into five categories: "Not Aligned," "Fully Aligned,"

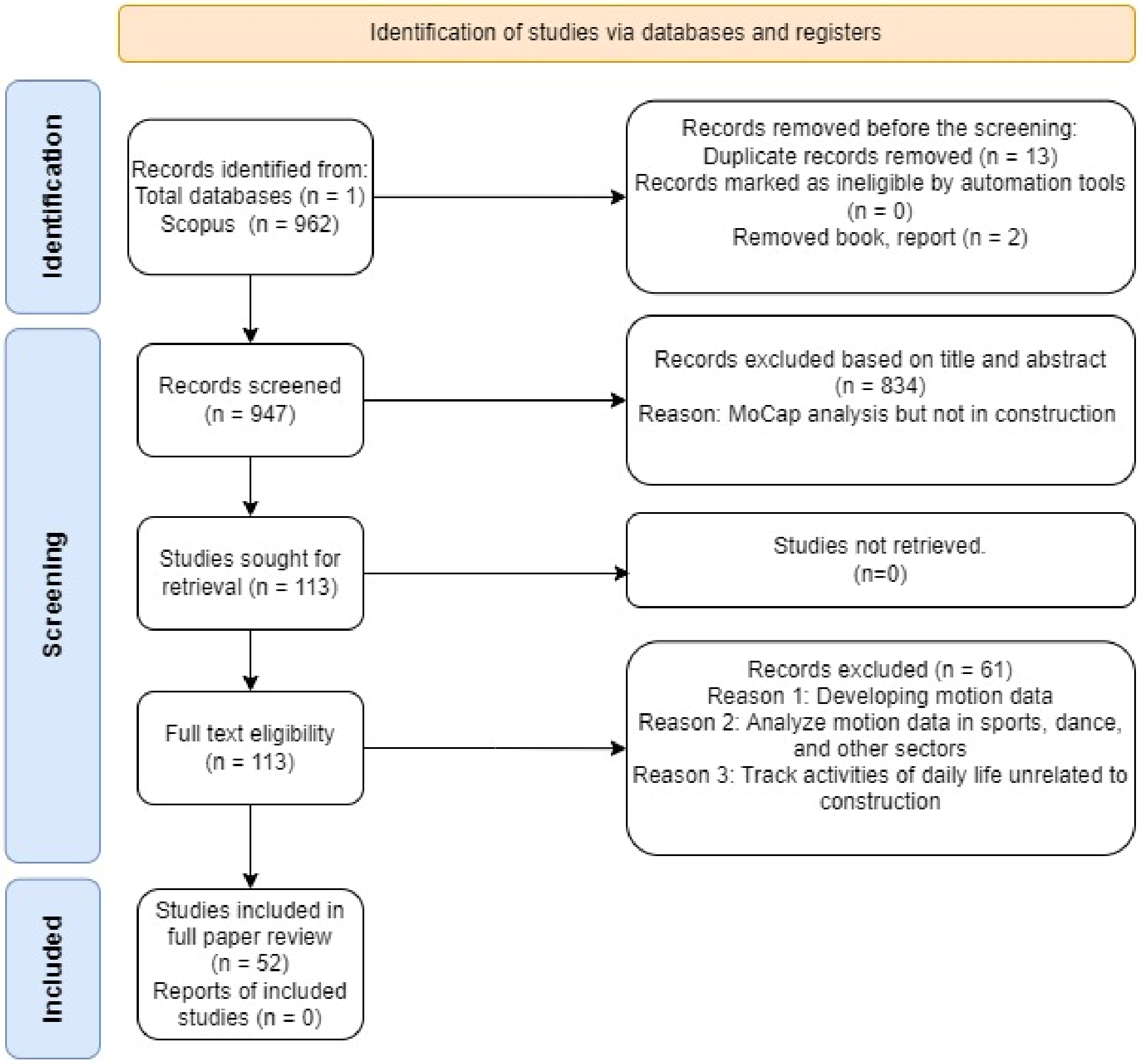

Figure. 2: Data extraction with PRISMA framework

"Partially Aligned," "Tangent Papers," and "Discussion of selection or rejection." Only publications classified as "Fully Aligned" by all three authors were included in our review. We excluded full-text articles for some reasons, such as the MoCap system being used to measure whole-body vibration for seated conditions, the technical aspects of motion data, and motion data configuration, which were outside of the scope of our study. Our study guarantees reproducibility and reduces biases by excluding articles that do not use MoCap systems or are unrelated to construction workplaces. This approach greatly enhances the reliability and precision of our comprehensive research, offering a more concentrated and precise overview of the utilization of MoCap systems in the construction industry.

# 3. Results

## 3.1. Yearly publication trends

Figure 3 provides an overview of research publication patterns on MoCap systems in construction workplaces from 1972 to March 2023. The bar graph shows that the field was in its early stages, with occasional activity and very few publications per year until a significant increase in publications began in 2002. There were also periods of time when no publications were produced, especially during the years 1974, 1975, 1977, 1978, 1982-1984, and 1992, which may suggest a lack of concentration or restricted technological capabilities during those decades. Starting in 2002, there was a progressive increase in publications, with a more noticeable surge from 2008 onwards. The years 2020, 2021, and 2022 had the highest number of publications, with 94, 83, and 83 documents, respectively, which indicates a recent surge influenced by several factors, including the COVID19 pandemic. The outbreak of the pandemic has accelerated the adoption of technology in various industries, including construction. As a result, there has been an increased interest in MoCap systems as a means of remote monitoring, training, and enforcing safety rules at construction workplaces. However, there was also a slight decline of about 11.7% in 2021 compared to 2020. Between 2008 and 2020, the number of publications rose from 25 to 94, representing a remarkable 276% growth, indicating a strong and expanding interest in the application of MoCap technology within the construction workplace. The growth can be attributed to advancements in wearable sensor technology, cost-effective price, availability of MoCap systems, and a stronger focus on using emerging technology to promote workplace safety and productivity.

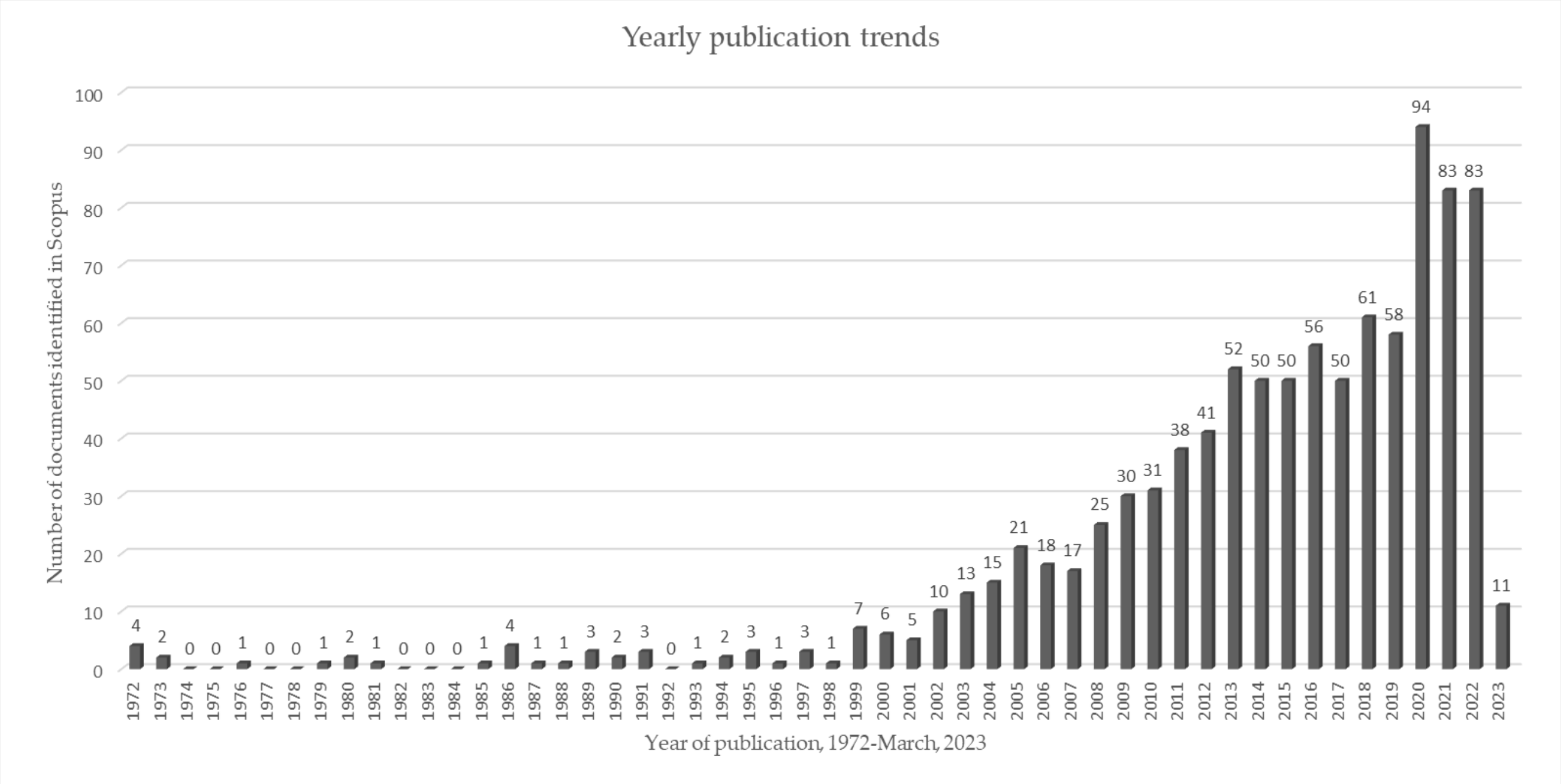

Figure. 3: Yearly publication trends (1972- March 2023)

## 3.2. Scientometric analyses

### 3.2.1. Keywords co-occurrence network analysis

The keyword co-occurrence network analysis is shown in Figure 4. For this analysis, in VOSviewer software, a minimum number of occurrences of a keyword was selected five, and a total of 15 keywords met the requirement among 565 keywords. Some identical terms were merged, such as workers, human, and humans merged as construction workers, motion capture, motion capture systems, and motion trackers merged as MoCap. In the network, the nodes (keywords) with larger circle sizes have a higher frequency of appearance in the literature, indicating their greater significance. The width of the links between nodes indicates the intensity of the association between keywords, with thicker lines indicating a more robust thematic connection. In the red cluster, *"MoCap"* is a prominent and significant node, which confirms its vital role in the research field, and it has strong associations with other notable entities such as *"occupational risks," "construction industry," "construction workers," "safety,"* and *"ergonomics."* The red cluster node suggests a significant focus on utilizing MoCap technology to improve safety, minimize injuries and ergonomic hazards, and enhance the physical well-being of workers in the construction workplace. Another notable cluster with green color is associated with *"WMSDs," "IMUs," "wearable sensors,"* and *"pattern recognition,"* indicating the utilization of wearable sensors to reduce WMSDs and the analysis of recorded motion data to identify patterns. On the other hand, the interrelationships among "biomechanics," "injury," and "WMSDs" underscore a health-focused study area that investigates the effects of MoCap technologies on the identification, tracking, and prevention of musculoskeletal problems for construction workers. The findings suggest that there is a strong acknowledgment of the possible advantages of MoCap technology in reducing occupational hazards and improving worker well-being.

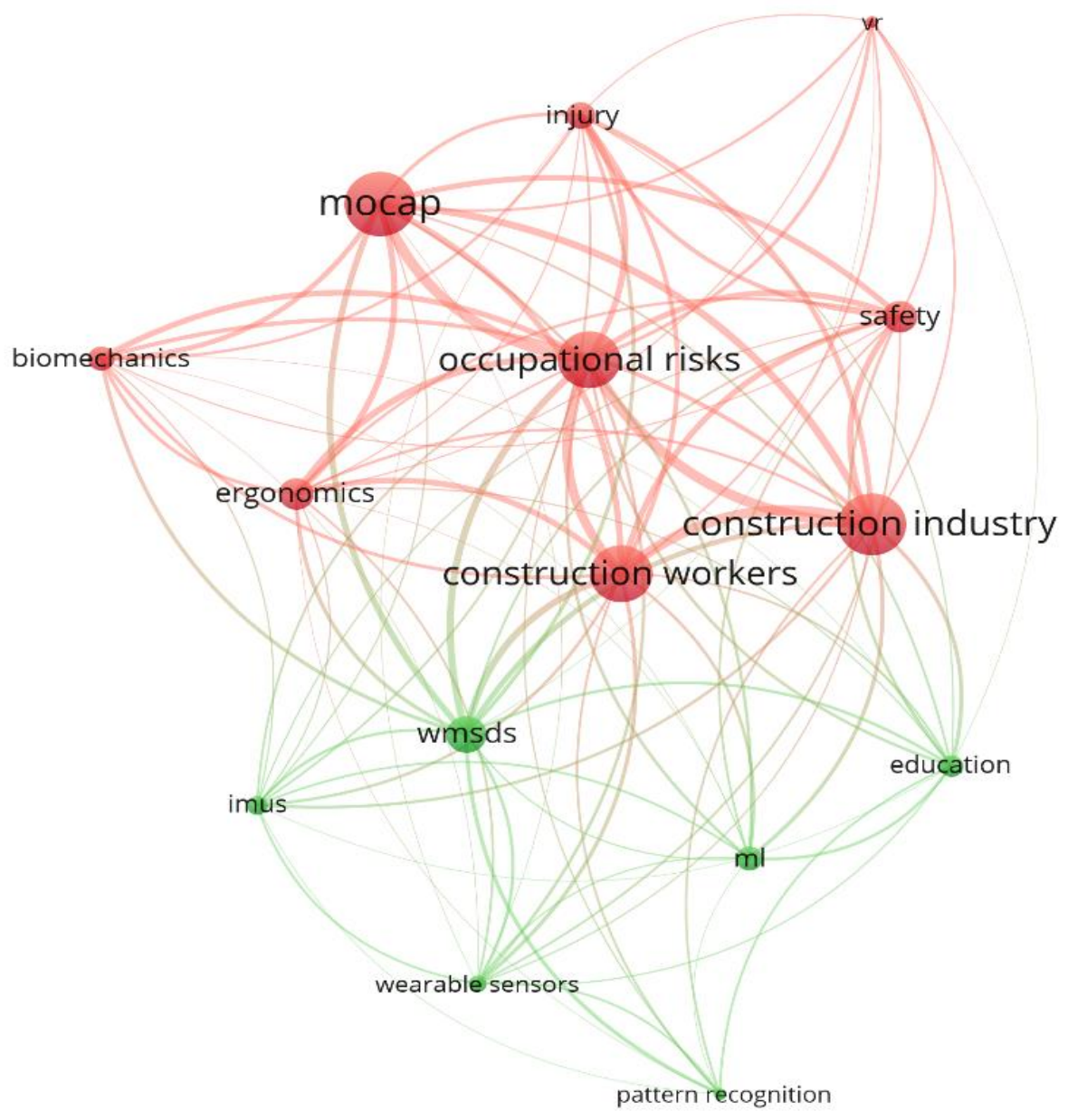

Figure. 4: Keywords co-occurrence network

### 3.2.2. Co-authorship by author analysis

The author co-authorship network analysis identifies five clusters, depicted by different colors, signify groups of researchers who frequently collaborate (Figure 5). The red cluster defined as highly significant, with "Han S." being the most productive and influential author, who has published eight papers and has made a significant impact in the academic community, as evidenced by the impressive number of 224 citations and the overall link strength of 11 indicates a wide range of collaborative connections. Nonetheless, the green cluster, "Abdel-Rahman E." has six documents and 48 citations, whereas "Haas C.T." and "Ryu J." each have five documents and a link strength of 11. On the other hand, "Li h." has five documents and 291 citations, as shown in the yellow cluster. Lastly, Obonyo e. and Zhao J. have six documents, 56 citation counts, and link strengths of 6, but they are independent nodes (violet cluster), suggesting new research directions or specialized contributions. The detail of the analysis is shown in Appendix A.

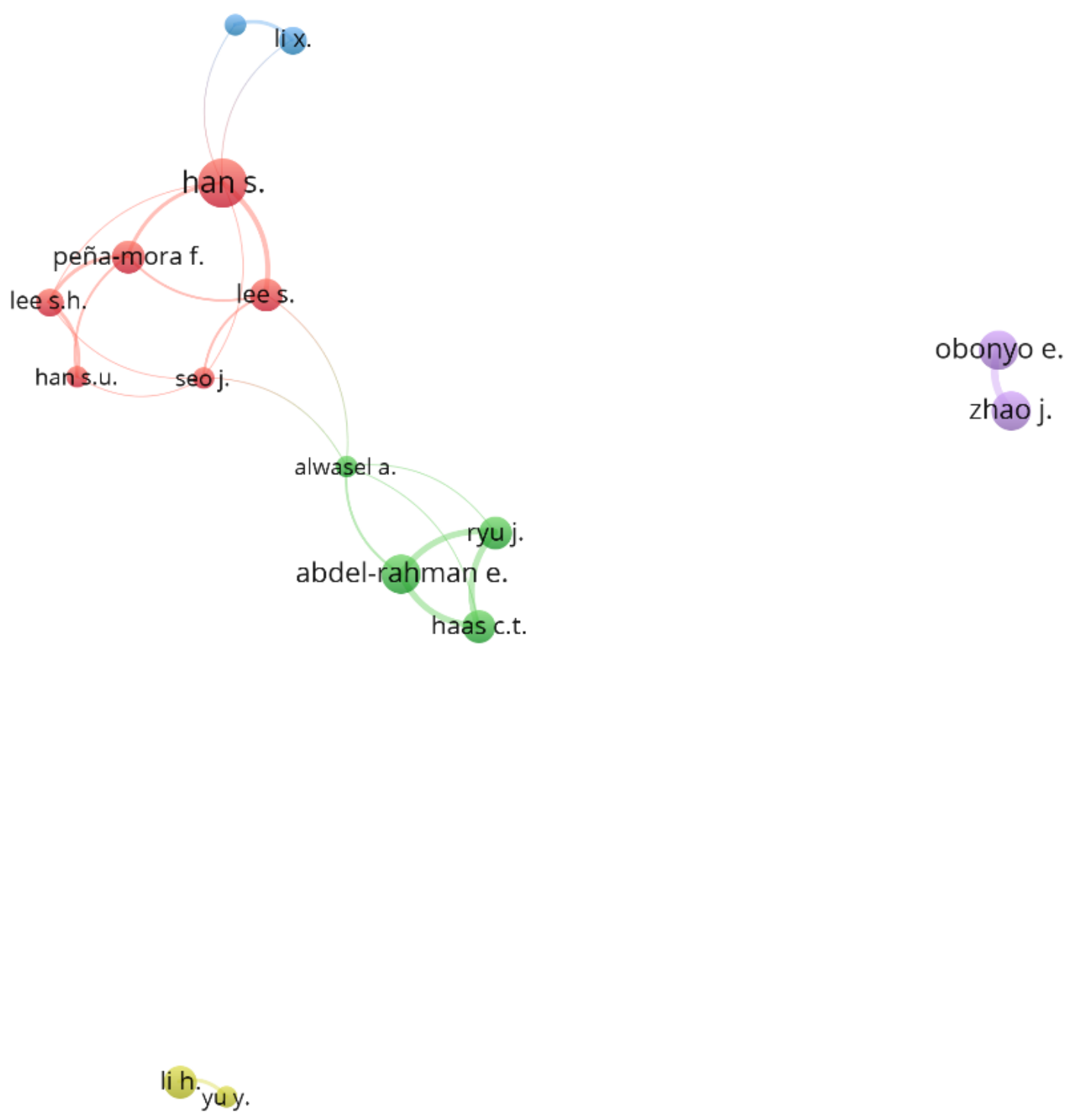

Figure. 5: Author co-authorship network analysis

### 3.2.3. Citation by document network analysis

The Citations by documents network analysis visually represents the scholarly influence and interconnectedness of publications as shown in Figure 6 (interconnected nodes are shown only), and Figure 7 shows the number of citations and links per document. The document authored by "Yin K. (2007)" has the highest level of influence with 242 citations, suggesting its significance and potential pioneering role in the field due to the novelty of the research, methodological benchmark, or relevance. Although lacking direct bibliographic ties to other publications in this analysis, the absence of links may indicate that "Yin K.'s" work is fundamental but may be unique in this area of focus or has had a significant impact on a wide range of studies that extends beyond the scope of the network being evaluated. Next in line is "Yan X. (2017)," with 186 citations and two links, showcasing its significant influence through the number of citations and its relevance within the network and suggesting that the document may have had a crucial role in shaping future research directions. Additionally, the documents authored by Valero E. (2016) and Chen J. (2017) have received significant attention, with citation counts of 136 and 116, respectively.

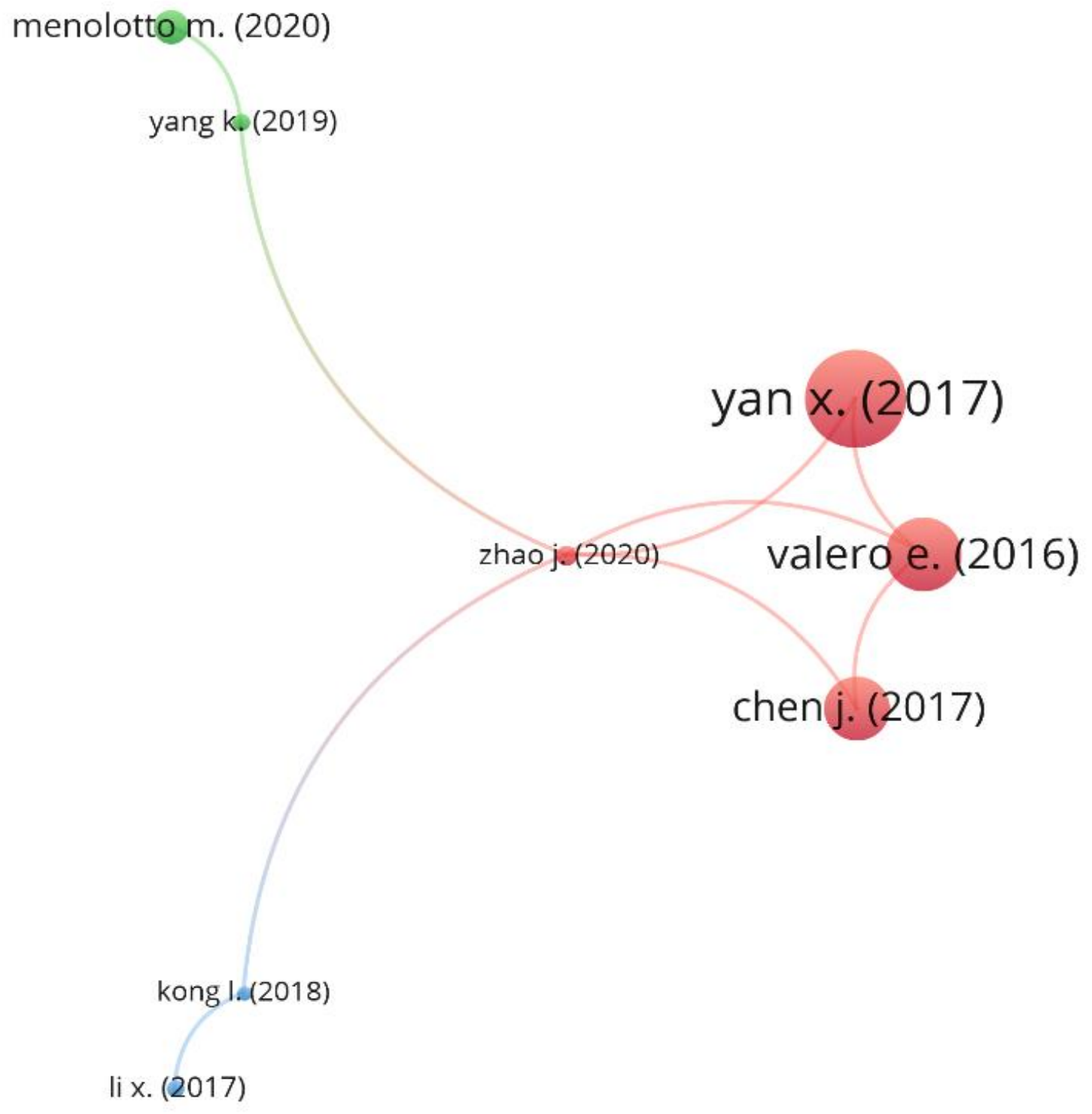

Figure. 6: Citation by document network analysis

Considering the clusters, the red cluster, with "Yan X. (2017)," is characterized by the most influential documents considering citation count and interconnectedness.

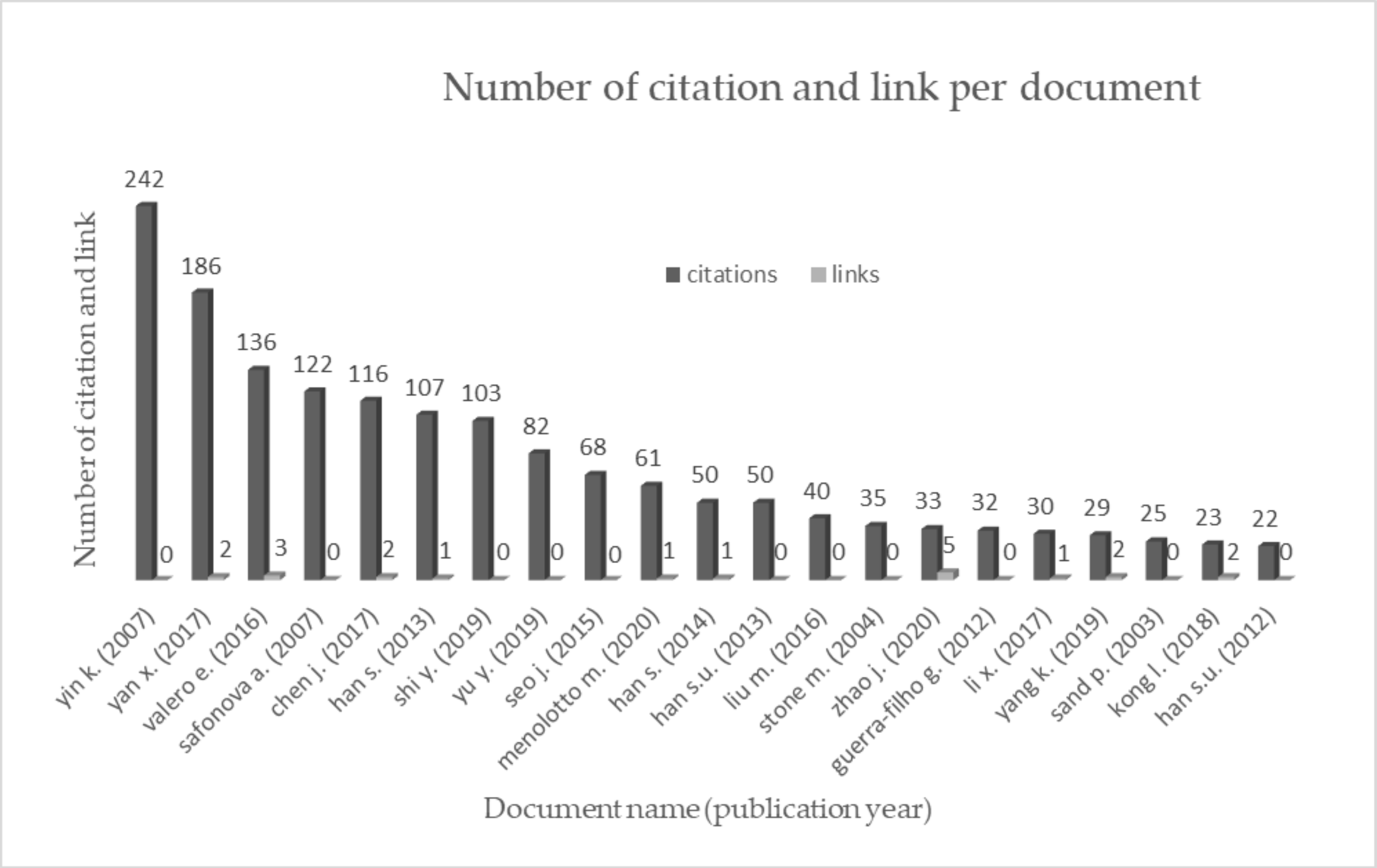

Figure. 7: Number of citation and link per document

### 3.2.4. Bibliometric coupling by countries

Bibliometric coupling by countries refers to when research papers from two nations reference the same articles; they are considered bibliometrically coupled. It helps visualize the collaborative links and intellectual connections among countries in a particular research domain. Table 1 shows that the United States has the highest number of documents (29) and citations (955), as well as the most robust total link strength (1120), indicating its prominent position in both productivity and influence in international collaborations. Canada has contributed significantly to the research network, with 16 papers receiving 456 citations and a strong link strength of 861. Moreover, Hong Kong has fewer documents (7). Still, it has a substantial citation count (426) and a strong link strength (794), indicating its influential research has been effectively incorporated into the global research network. Similarly, China has also actively engaged in the research network, with a document count of 7, a citation number of 410, and a significant link strength of 633. Figure 8 depicts the visualization network where the red cluster comprises the USA, Canada, India, and South Korea, which share similarities in terms of their research references, research methodologies, and thematic emphases in MoCap technology. On the other hand, the blue cluster, which includes Hong Kong and China, signifies a distinct collection of common points of reference or research topics within these nations. This difference may arise from various sources, such as regional research agendas, specific technical breakthroughs, or collaboration patterns within these countries.

Table 1. Bibliometric coupling by countries

| Country | Documents | Citations | Total Link Strength |
|---|---|---|---|
| United States | 29 | 955 | 1120 |
| Canada | 16 | 456 | 861 |
| China | 7 | 410 | 633 |
| Hong Kong | 7 | 426 | 794 |
| India | 2 | 0 | 12 |
| South Korea | 2 | 33 | 156 |

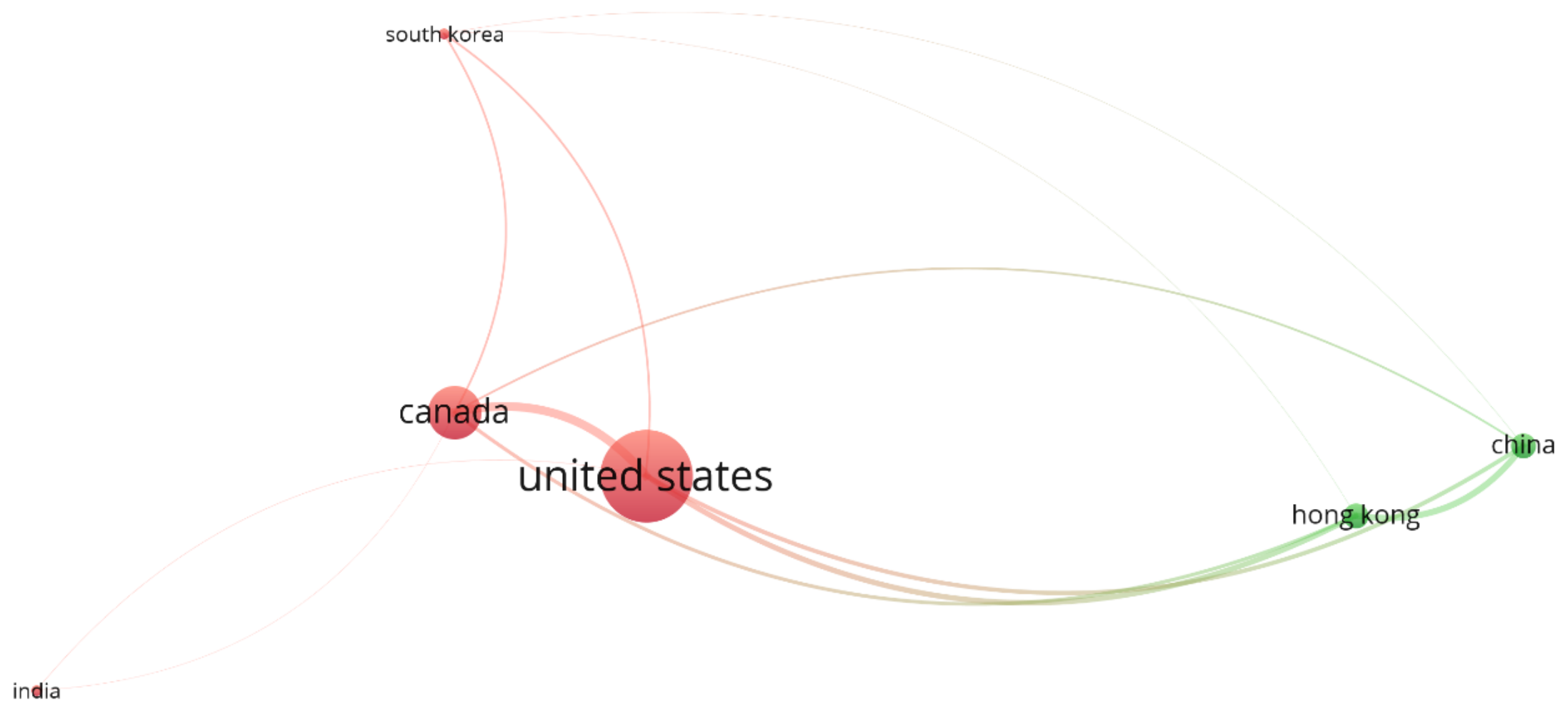

Figure. 8: Bibliometric coupling by countries network analysis

## 3.3. In-depth analysis

### 3.3.1. Types of construction work

After analyzing 52 articles on various types of construction work, it is clear that most of the research focuses on general construction tasks like lifting, climbing, or training, which make up 48% of the publications. Masonry is also a significant construction task for research, addressed in 20% of the publications. The activity of ladder climbing comprises 12% of the total reported studies. Still, specialist jobs such as bricklaying/concreting, carpentry and painting, rebar, building roofing, and inventory and structural work have lower representation, ranging from 2% to 6%. The distribution of jobs in MoCap research indicates a preference for basic construction tasks for research purposes, while other areas have a more specialized or unique use of MoCap technology.

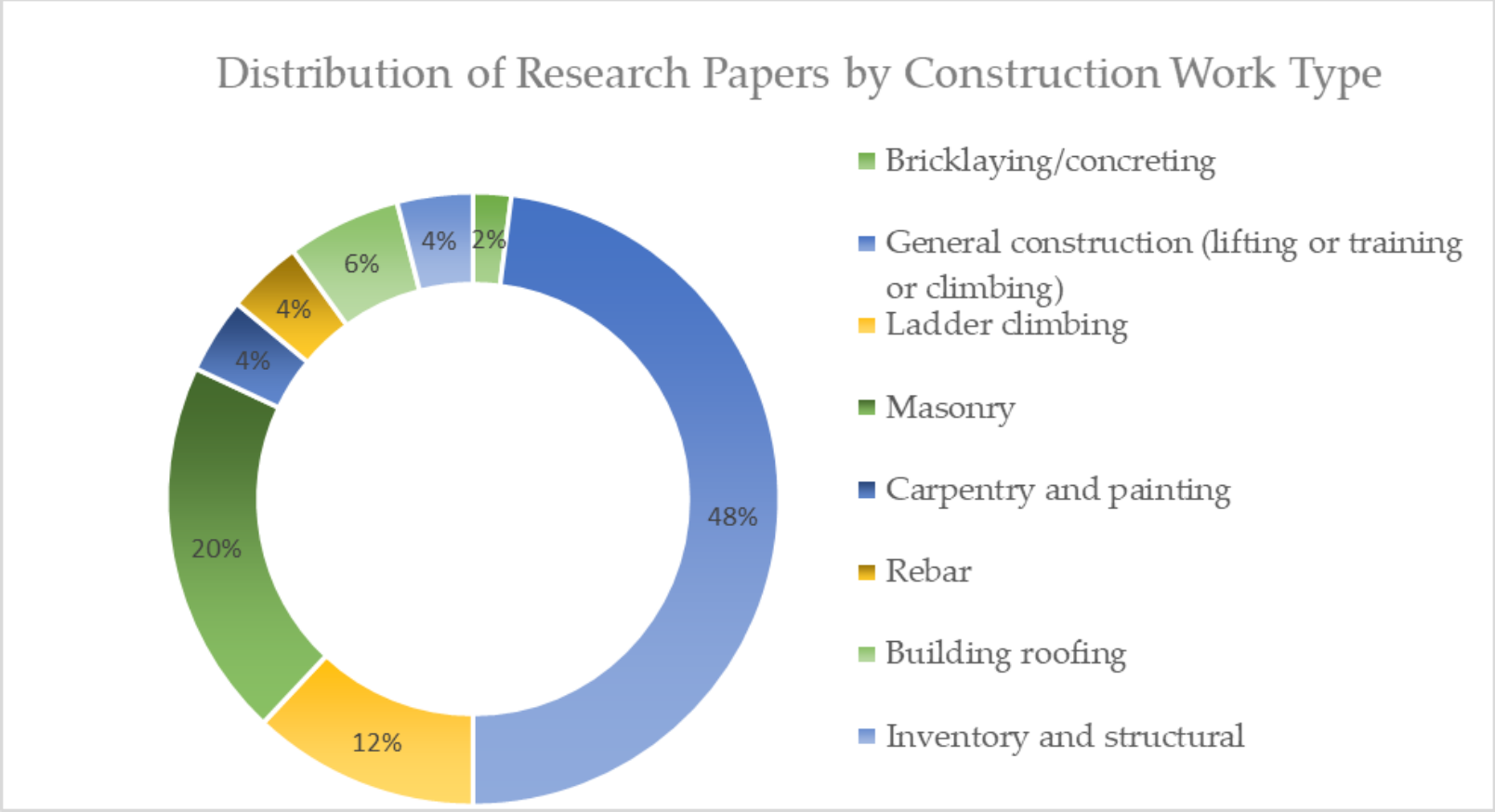

Figure. 9: Distribution of research papers by construction work type

### 3.3.2. Type of MoCap technology

The summary of the type of MoCap technology used in reported studies is shown in Table 2. It is evident that "Vision-based and camera systems" are the primary focus, accounting for roughly 37% of the studies (19 papers). These technologies, such as Microsoft Kinect, video cameras, or RGB-D sensors, are clearly inclined toward visual motion tracking in research. The use of IMUs is about 28.85% of the reported studies (15 papers), which indicates a significant dependence on this technology. Hybrid technologies, which involve IMUs with VR, Kinect, or smartphones, together with specialized systems such as the Noitom Perception Neuron, make up about 11.54% of the studies (6 publications). These technologies demonstrate an integrated approach to MoCap systems in the era of Industry 5.0. The importance of VICON systems and other vision-based technologies is evident in 13.46% of cases, as indicated by seven studies. Finally, a small but increasing number of investigations (4 publications, accounting for 7.69% of the total) employ advanced technologies like EMG and VR for modeling, indicating a developing interest and new research trends in these current approaches. The findings demonstrate the widespread adoption of vision-based and camera systems in the construction industry, indicating a shift towards technologies that provide user-friendly interfaces and comprehensive data collection. The prominent use of IMUs and hybrid systems shows a versatile methodology for MoCap, amalgamating the advantages of various technologies.

Table 2. Type of MoCap technology used in reported studies.

| MoCap technology type | Technology included | Number of papers | Studies | Percentage of studies (%) |
|---|---|---|---|---|
| IMUs | 17, 5,6,7 IMUs were used | 15 | [61–75] | 28.85 |
| Hybrid systems | IMUs and VR, IMUs and Kinect, IMUs and smartphone, Noiton perception neuron | 6 | [76–81] | 11.54 |
| Vision-based and camera systems | Microsoft Kinect, video camera, RGB-D, smartphone, motion camera | 19 | [82–100] | 36.54 |
| Optical MoCap | VICON, vision-based, direct optical encoder system | 7 | [91,101–106] | 13.46 |
| Contemporary technology | EMG, VR for simulation, helmet mounted | 4 | [107–110] | 7.69 |

On the other hand, considering the use of markers (Table 3), it was found that the marker-less systems were the most frequently used method, making up 67.31% of all reported studies, suggesting that non-invasive motion tracking approaches are the preferred choice for current research. Nonetheless, marker-based systems were mentioned in 9.62% of the publications, which indicates that they are still valuable but less commonly used than marker-less technologies. A small percentage, precisely 5.77%, investigated a combination of marker-less and marker-based methods, showing an interest in hybrid systems that can utilize both benefits. Around 17.31% of papers did not provide specific information about the type of system used.

Table 3. MoCap technology based on the use of markers.

| Types | Number of papers | Studies | Percentage of paper |
|---|---|---|---|
| marker-based | 5 | [87,102,104,105,108] | 9.62 |
| marker-less | 35 | [61–63,67–75,77–82,84–86,88,90,91,93–95,97–101,106,110,111] | 67.31 |
| Combination of marker-less and marker based | 3 | [76,78,103] | 5.77 |

### 3.3.3. Sensor attachment location

The sensors were placed on various areas of the human body to capture motion, where approximately 9.62% of the studies focused on the upper limb (including the head), but only 3.85% focused on the lower limb (Table 4). Over half of the studies (53.85%) aimed to capture the entire body's motion, while only 7.69% focused on the hand. Nonetheless, around 25.00% of the studies didn't mention the specific areas where the sensors were attached. Tracking motion from the whole body allows for a thorough understanding of human biomechanics and ergonomics, which is particularly important in the construction industry, where complex activities and movements occur. It provides a holistic view of human motion, crucial for accurate ergonomic assessments, injury prevention, and increased efficiency. The approach helps understand the relationships between various anatomical components better, providing valuable insights for developing safer and more efficient work environments.

Table 4. Sensor attached locations on human body.

| Sensor attached location | Number of papers | Studies | Percentage of study |
|---|---|---|---|
| Upper limb (including head) | 5 | [66,79,83,107,110] | 9.62 |
| Lower limb | 2 | [88,106] | 3.85 |
| Whole body | 28 | [61–65,67–72,74–77,80,81,84,87,90,94,98,99,102–105,108] | 53.85 |
| Hand | 4 | [73,78,89,109] | 7.69 |

### 3.3.4. Theme analysis

Theme analysis with the selected papers shows that the majority of the papers studied, comprising 38.46%, emphasized "technology and design improvement," as shown in Table 5, which highlights the industry's strong inclination towards developing new technology, increasing accuracy (%) of the developed method and improving design approaches. In contrast, only 7.69% of the research works were related to "performance optimization," which suggests that although it is a crucial component of the construction business, it has not received the same scholarly attention and implies a promising area for future research. In addition, the literature extensively covers the themes of "safety and risk management" and "health and ergonomics," which account for 26.92% and 25.00% of the publications, respectively. The findings indicate that the construction industry is highly dedicated to innovation, particularly technology and design. However, there seems to be a lack of emphasis on performance optimization, which presents a promising opportunity for future research. Additionally, the industry's strong focus on safety, risk management, health, and ergonomics reflects its commitment to protecting workers and minimizing potential ergonomic hazards. As a result, the construction appears to have a well-rounded workplace that prioritizes technological advancement while ensuring safety and efficiency.

Table 5. Themes of reported studies

| Themes | Number of papers | Studies | Percentage of paper |
|---|---|---|---|
| Performance optimization | 4 | [70,82,89,105] | 7.69 |
| Safety and risk management | 14 | [61,63,64,64,65,69,71,74,78,79,87,88,99,104] | 26.92 |
| Health and ergonomics | 13 | [66,72,73,75,81,83,84,98,100–102,107,108] | 25.00 |
| Technology and design improvement | 20 | [62,67,68,76,80,85,86,90–97,103,106,109–111] | 38.46 |

### 3.3.5. Summary of study designs, application, and recommendations

The summary of study designs, applications, and recommendations are illustrated in Table 6. We found a wide range of sample sizes, varying from 1 to 126 individuals, which indicates significant differences in research scale and objectives. Researchers have employed various sensors and technologies, such as Kinect sensors, smart insoles, IMUs, and cameras, to examine movement in a multidimensional manner. Focusing on MoCap applications, the systems are widely used in the construction industry for various purposes, such as estimating labor productivity, analyzing work duration, monitoring ergonomic hazards, and preventing fall risks. Additionally, designing worksites according to Mechanical Energy Expenditure (MEE), identifying ergonomically hazardous postures, evaluating worker tiredness in real-time, and instructing apprentices or novices to avoid risky postures are some other applications. It also enables behavior monitoring, demonstrating their versatility in enhancing the construction industry's safety, health, and efficiency.

On the other hand, the recommendations emphasized improved accuracy in estimating productivity, the use of non-invasive measurement methods, real-time assessment, and the possibility of 3D visualization-based modeling to mitigate ergonomic hazards. A noticeable trend is the increasing use of advanced MoCap technologies, especially in virtual reality (VR) settings, indicating the field's progression towards using technology to improve safety, performance, and reduce ergonomic hazards in construction workplaces. The growing reliance on emerging technologies for precise motion analysis demonstrates an ever-increasing dependence on technology for thorough ergonomic evaluations.

Table 6. Summary of study designs, applications, and recommendations.

| Study | Sample size | Number of sensor or devices | Applications | Recommendation/innovation |
|---|---|---|---|---|
| [82] | 5 | 1 Kinect sensor and 1 depth camera | Labor productivity estimation and compare labor cost and duration of task can be assessed. | Reliable productivity estimates can be achieved without huge historical data. |
| [101] | 1 | 2 smart insoles | Mechanical Energy Expenditure (MEE) is used to design worksite. | This is a non-invasive process to measure MEE. |
| [61] | 6 | 17 IMUs | Jerk can be utilized to calculate physical exertion and fatigue. | Feature selection method to identify motion changes, optimal sensor number and placement could be considered to improve the method. |
| [62] | 4 | 5 IMUs | Model test for awkward posture recognition. | - |
| [102] | 3 | - | Ergonomically risky postures can be recognized. | A 3D visualization-based modeling method is suggested to reduce ergonomic risks. |
| [83] | 126 | - | Effect of social influence on construction workers' behavior can be assessed. | The study only focused on high rise building hazardous situation, and calibration accuracy between actual movement and virtual animation could be improved. |
| [63] | 45 | 17 IMUs | Apprentice training to avoid risky posture and improve movement. | Biomechanical analysis can be used to determine joint forces and moment, and efficiency and productivity can be analyzed. |
| [76] | 4 | 3 cameras, 2 IMUs | Ergonomic hazard monitoring and fall risk prevention. | Further study can be conducted on gait analysis-based monitoring technique for the construction environment. |
| [107] | 126 | 1 camera | Interpersonal influence of workers' mistake and unsafe behaviors. | More accurate motion tracking system could be used, other hazardous scenarios could be investigated. |
| [77] | 4 | VR and IMUs | - | - |
| [64] | 1 | 5 IMU | Proper placement of sensors, increase computational performance. | More participants can be included to validate the proposed model, assess long term operation, and real time performance for activity recognition. |
| [65] | 45 | 17 IMUs | Expert and apprentice work diff, training, minimize health and safety risks. | Other kinematic factors such as joint angle, carrying distance, joint loads may affect safety and productivity |
| [103] | 1 | 2 | MoCap system can be used to identify potential safety and health risk. | - |
| [66] | 3 | Virtual Reality environment | Construction safety training through virtual reality. | First time VR users will be collected for holistic approach, and apprentices could be incorporated. |
| [67] | 4 | 5 IMUs | Monitor risky or injury related postures and prevent them. | The developed model will be deployed on mobile device for real-time safety. |
| [84] | 4 | 13 UMUs | Fatigue assessment in real-time automatically and invasively. | The mass of material and tools should be considered in the future studies and workers rest status could measure automatically. |
| [85] | 1 | 1 Kinect | Detect motion data in real-time, automatically process biomechanical analysis to detect the risk of WMSDs. | The reliability of this study needs to be identified. |
| [86] | 1 | 1 Kinect | Understanding the causes of WMSDs in manual construction work. | Marker-based MoCap system can be used. |
| [87] | 1 | 8 cameras | Identify unsafe actions while lifting heavy objects and increase productivity by reducing work dimension. | Marker-less system and sufficient data will be used in the future. |
| [88] | 2 | 1 Kinect | Prevent WMSD related to ladder climbing and identify causes of falls | However, the process is less accurate than marker based MoCap system. In future, how error in motion data affects joint moments can be identified. |
| [78] | 1 | 2 (1 Kinect and 1 IMU) | Combination of two system could enhance the safety measures in construction. | In future, Kinect data can be used to improve accuracy of IMUs through data fusion. |
| [89] | 2 | 2 smartphones | To increase productivity, analyze workflow processes and remove bottlenecks. | - |
| [79] | 1 | 2 IMUs and 1 smartphone | Real-time hazard and WMSDs prevention in construction. | Solar-charged or chargeless IMUs could be employed to monitor real-time ergonomic hazards. |
| [90] | 1 | 1 Stereo MoCap and 1 Kinect (RGB-D) | The proposed stereo vision cameras can be used for high data precision and environmentally harsh condition. | Pose estimation techniques can be expanded and more harsh environmental condition will be tested measuring motion data using multiple cameras. |
| [91] | 1 | 1 Kinect, and iPi soft MoCap | Detect unsafe actions using depth sensor. | The developed system is not applicable in outdoor construction sites. Light condition, effect of occlusion and enough sampling could be considered. |
| [92] | 1 | 1 smartphone | Hazard prevention and alert system are studied. | Multiple hazards (stationary and moving) can be considered, and more complex settings can be tested. |
| [68] | 4 | IMUs | Compress and reorganize motion data and use less power and memory while collecting and analyzing motion data. | Number of tensing channels can be reduced; accuracy can be increased, and data fusion and distributed data processing can be included. |
| [93] | 1 | 2 smartphone cameras | Behavior monitoring, ergonomic assessment, and productivity analysis. No sensor was attached to human body. | 3D construction accuracy could be studied later. This data can be used for ergonomic assessment and productivity analysis. |
| [94] | 1 | 1 Kinect | Inaccurate motion data and errors should be corrected for ergonomic analysis. | Real error of motion data will be used instead of random error data in the future. |
| [95] | 1 | 1 Kinect | Unsafe action recognition using motion data (rotation angle, joint angle, position vector). | Large field data with more movements (lifting, slipping, twisting kneeling) will be measured. For real time monitoring, kernel PCA can be applied. |
| [96] | 1 | 1 Kinect and 1 Vicon MoCap | Accuracy of Kinect and VICON can be measured. | Kinect system may not be applicable for hand related ergonomic analysis. Other actions like walking, running, lifting and carrying object can be evaluated by it. |
| [111] | 1 | 1 Computer vision (camera) | The approach can detect unsafe actions using motion data. | Future study will evaluate the validity of this approach for automatic unsafe actions detection. |
| [108] | 3 | sEMG sensors | Muscle fatigue can be identified following the study. | This model can be used to minimize muscle forces with the final goal of preventing muscle injuries including more experimental data. |
| [97] | 1 | 1 Kinect | L5/S1 disc, left knee and elbow were considered for data collection. | Reliability and accuracy of motion data, BVH, needs to be improved with collection data from more joints. |
| [69] | 8 | IMUs | Reduce WMSDs and increase safety, health, and productivity. | Working height should be reduced to minimize WMSDs and waist level could be optimal height for individual material handling. |
| [109] | 10 | 1 armband and 1 metabolic analyzer | The proposed system is suitable for construction workers continuous fatigue measurement to assess physiological status and early detection of risk. | Only forearm band was considered for data collection whereas other joints could be considered to measure physical fatigue. |
| [81] | 10 | IMUs | Whole body joints were analyzed. | Integrated CML data can be used for further analysis. |
| [73] | | 1 | Established that the accuracy of the action recognition with raw dataset. | Acceleration of forearm was analyzed, but other body parts could be considered. |
| [80] | 5 | 17 | Improvement of workstation design through reduction of prototype iteration and time. | - |
| [98] | 8 | 3 | Posture recognition using smartphone. | In pocket, smartphone can be used to recognize posture. |
| [72] | 7 | 5 | Generative Adversarial Network in conjunction with DNN can improve the posture recognition. | - |
| [100] | | 1 | Objective approach is better for fatigue detection. | Quantitative assessment could be incorporated. |
| [74] | 30 | 5 | Designed wearable sensing system to detect posture and improve worker safety awareness. The UI can deliver actionable MSD risk assessment to user readily. | Workers and manager real-time feedback systems could be studied to reduce WMSDs. |
| [70] | 66 | 17 | Getting experienced workers' postures, it can be applicable for apprentice training program. | Bricklaying task was considered, however, other tasks such as climbing, welding, and cutting could be considered. |
| [75] | 1 excavator | 6 | The study can be applied for real-time productivity monitoring, safety analysis, and dynamic simulation input in construction. | Focus on more generalizable and accurate activity identification models, potentially enhancing automated monitoring and operational analysis in construction environments. |
| [110] | 1 | 1 | The system can be used to improve the forward leaning and squatting postures. | Future studies could incorporate more construction workers. |
| [104] | 9 | 1 | Reveals that dynamic kneeling postures during shingle installation can generate significantly higher knee flexion, abduction, adduction, internal and external rotation compared to static postures, indicating increased MSD risk. | More professional roofers, assessing work-related factors like different slopes and postures on knee MSDs, observing EMG signals of knee muscles and joint contact stress. |
| [99] | - | 1 | Using smartphones equipped with accelerometers and gyroscopes can detect and identify near-miss falls in construction workers. | Enhance algorithm accuracy and apply in real-world construction sites. |
| [71] | - | 7 | - | Novel wearable wireless system can be used to long term and ubiquitous tracking of body posture and motion in noninvasive way. |
| [105] | 1 | 8 | Highlights the feasibility of using the OpenSim biomechanical analysis tool and presents a case study on ladder climbing activities using motion data from VICON. | Marker-less motion data can be applied for assessing WMSDs. |
| [106] | 1 | 1 | Knee joint angle measurement without complex mathematical operations. | - |

## 4. Discussion

The results of the comprehensive analysis of MoCap technology in construction are crucial within the framework of Industry 4.0 and 5.0 [112,113]. Integrating digital transformation and human-centered innovation improves efficiency and safety in the construction workplace [114–116]. The study emphasizes the importance of the use of MoCap systems to assess and enhance ergonomic safety and worker well-being as part of the transition to Industry 5.0, aiming to prioritize human-centric practices and align human skills with technical advancements. The annual publication trends reveal a growing interest in MoCap technology in the construction workplaces, aligning with the digital upsurge in Industry 4.0. The scientometric analysis demonstrates that the industry places significant emphasis on occupational risks, safety, and ergonomics, highlighting its dedication to promoting worker health, a fundamental principle of Industry 5.0. The examination of the keyword co-occurrence network demonstrates a significant emphasis on occupational hazards, safety measures, and ergonomics in the field, highlighting the industry's commitment to ensuring the well-being of its workers [117,118]. The prevalence of terminology such as "IMUs" and "wearable sensors" in this network highlights the technical transition towards more sophisticated, non-intrusive, real-time, automated, and all-encompassing MoCap techniques [119,120]. Furthermore, the author co-authorship network analysis highlights how researchers collaborate with each other, taking a multidisciplinary approach in this field. The citation by document analysis identifies significant studies that have had a powerful impact and demonstrates a vigorous exchange of knowledge and fundamental research that has shaped the development in this field. Finally, the bibliometric coupling by country analysis reveals the global scope of research, with countries such as the United States and Canada leading the way in making contributions. The widespread connectivity of the world illustrates the broad usefulness and importance of MoCap technology in the construction workplace [1]. It transcends geographical limits, promotes international collaboration and indicates a vibrant and interconnected research environment driven by collaboration, significant studies, and global involvement.

The in-depth review reveals a diverse variety of sample sizes, ranging from individual case studies to larger cohort analyses, which not only showcases the distinct research methodologies but also demonstrates the versatility of MoCap technologies in numerous construction-related situations. Moreover, the primary focus of the construction tasks examined primarily encompasses fundamental actions such as lifting, training, and climbing, which are essential components of daily operations inside a construction workplace [121]. The emphasis on these jobs highlights the dedication of researchers to enhancing the functional aspects of work and guaranteeing the ergonomic well-being of these regularly executed motions. The extensive utilization of emerging technologies such as Kinect sensors, IMUs, and VR technologies indicates an industry that is leading in incorporating state-of-the-art technology for practical purposes [122–124]. This trend suggests a movement towards non-invasive and user-friendly technologies, while also providing extensive data collection capabilities in real-time, which are essential for improving safety and on-site productivity [125,126]. In addition, the research indicates a preference for marker-less systems, which are more streamlined, less obtrusive and non-invasive for capturing human motion. The strategy is in line with the requirement for pragmatic, tangible solutions that minimize interruptions to the organic workflow in the construction workplaces [127,128].

The findings of this systematic analysis emphasize a changing profession that is increasingly dependent on advanced technology to tackle conventional obstacles in the construction workplaces. The industry's emphasis on worker safety, ergonomic enhancements, and the implementation of sophisticated MoCap systems demonstrates its active embrace of technical advancements and recognition of their capacity to transform conventional methods. The findings obtained from this review provide valuable information about current trends and practices and establish a foundation for future research endeavors focused on improving worker safety and operational efficiency in the construction workplace.

## 5. Future research directions

Research into MoCap technologies has revealed certain limitations that need to be addressed in future studies. There is a lack of multimodal motion assessment techniques, such as assessing two or more motion parameters (joint angles, acceleration, jerk, spatial position, and orientation), which are necessary for conducting thorough safety assessments [129,130]. In addition, the lack of real-time and automated data collection makes monitoring and analyzing movements when they occur efficiently difficult. Additionally, limitations include quickly identifying and correcting risky or potentially hazardous body positions. To address these issues, real-time feedback systems in tactile, visual, or auditory feedback could be incorporated to promptly notify users of potential risks and guide them to take corrective actions. Improving these elements makes MoCap systems more effective and useful in real-time industrial applications. Another major challenge is obtaining precise three-dimensional kinematics without markers, as in traditional systems [131]. Machine learning techniques promise to overcome this issue, but their suitability for biomechanical applications needs further exploration [132]. Additionally, most validation studies focus on slow or single-plane movements, emphasizing the need to thoroughly evaluate sport-specific, faster movements. Moreover, conventional markers may lead to inaccuracies when used simultaneously with marker-less MoCap systems. Overcoming these limitations can significantly enhance the accuracy and usefulness of MoCap systems in various applications.

To further enhance the applicability of MoCap technology in the construction industry, future investigations should prioritize utilizing real-time data analysis to make informed decisions on worker safety and productivity. Integrating MoCap technology with Industry 4.0 technologies such as AI and IoT (Internet of Things) can create highly responsive systems that can accurately detect risky postures, provide feedback, and guide to reduce ergonomic hazards. In addition, to ensure broader industry acceptance, MoCap systems should be customizable, portable, and cost-effective to different construction conditions and workforce profiles while being pleasant, unobtrusive, less invasive or non-invasive, and user-centric and should be capable of monitoring real-time health and ensure the safety of construction workers [133,134]. A single smartphone software or less invasive sensors such as Wristband could be studied and utilized to detect ergonomic hazards in real-time, provide feedback to workers and recommend necessary actions to avoid injury in construction workplaces. Furthermore, it is also crucial to address privacy and ethical concerns as MoCap technology becomes more integrated into construction operations. On the other hand, promoting interdisciplinary collaboration can foster the creation of comprehensive solutions that are both technologically advanced and practically feasible in the construction workplaces. Moreover, the review paper has certain limitations due to its reliance on the Scopus database, which might restrict the range of available material. While ensuring a focused and consistent analysis was effective, including multiple databases, such as Web of Science, PubMed, ScienceDirect, and Google Scholar, could have provided more viewpoints. Although it ensured consistency, using the same number of papers (n = 52) for both scientometric and in-depth analyses may have limited a broader perspective of the topic. More studies could be considered for scientometric visualization analysis in the future research. These suggestions and areas of study indicate future research recommendations in which MoCap technology could significantly contribute to improving health, safety, and productivity in the construction workplace.

## 6. Conclusion

The study provides a comprehensive review of MoCap systems in the construction workplace, set against the backdrop of Industry 4.0 and the emerging Industry 5.0 paradigms. MoCap technology is increasingly used in construction to enhance safety and improve ergonomic practices. The study analyzed 52 selected studies and conducted bibliometric, scientometric, and in-depth systematic reviews. The scientometric results showed that the high frequency of keywords such as *"MoCap," "occupational risks," "construction workers," "construction industry,"* and *"WMSDs"* indicate that the research community is heavily focused on using MoCap technology to understand and mitigate the occupational hazards in the construction industry and increasing awareness of the health and safety challenges faced by construction workers, specifically concerning Work-Related Musculoskeletal Disorders (WMSDs). In addition, the United States is a leading country in MoCap research, contributing a significant number of documents (29), citations (955), and total link strength (1120), which indicates that the United States is responsible for a considerable portion of fundamental and influential research in MoCap technology, particularly in construction. Additionally, it suggests that the nation has strong academic and research networks, which contribute to the global repository of information in this field.

On the other hand, the in-depth analysis found that lifting, climbing, training, and masonry are the most assessed construction tasks utilizing MoCap technology, primarily through marker less and IMU sensors. Most researchers (53.85%) considered whole-body motion tracking analysis. It indicates the need for a comprehensive analysis, essential for understanding complex activities and actions through examination of human biomechanics and ergonomics, which is crucial for accurate evaluations, injury prevention, and improved efficiency. By analyzing whole-body movements, researchers can gain a better understanding of how various body parts interact with each other, providing useful insights for the development of work environments that are both safer and more efficient. The review emphasizes the themes of selected studies and found that 38.46% of studies focused on “technology and design improvement,” around 23% emphasized “safety, and risk management,” and 25% considered “health and ergonomics.” By prioritizing “safety and risk management,” and “health and ergonomics,” it shows the commitment to protecting worker well-being and minimizing ergonomic hazards, which also aligned with the results of the scientometric analysis.

Finally, the study suggests areas for future research, including enhanced real-time data processing, integration with Industry 4.0 and 5.0 technologies, customization for diverse construction environments, a focus on worker-centric design, and addressing privacy and ethical considerations. The review paper contributes significantly to the body of knowledge in the field of technology use in construction, highlighting the potential of MoCap technology to revolutionize the construction industry by enhancing safety, efficiency, and overall worker well-being.

**Author Contributions:** Conceptualization, M.H.R., M.R.H.; methodology, M.H.R. and M.R.H; software, M.H.R.; validation, M.H.R. and N.I.C; formal analysis, M.H.R., N.I.C; investigation, M.H.R.,N.I.C,M.R.H and M.U.F; data curation, M.H.R.; writing—original draft preparation, M.H.R.; writing—review and editing, M.H.R., M.R.H, M.U.F and M.A.B.S; visualization, M.H.R.; project administration, M.H.R and M.R.H

**Funding:** This research received no external fundings.

**Conflicts of Interest:** The authors declare that there are no conflicts of interest or personal connections that could be perceived as influencing the research presented in this paper.

## Appendix A

Appendix A. Co-authorship by author network analysis

| Author | Documents | Citations | Total Link Strength |
|---|---|---|---|
| Han S. | 8 | 224 | 11 |
| Abdel-Rahman E. | 6 | 48 | 12 |
| Obonyo E. | 6 | 56 | 6 |
| Zhao J. | 6 | 56 | 6 |
| Haas C.T. | 5 | 37 | 11 |
| Lee S. | 5 | 129 | 9 |
| Li H. | 5 | 291 | 3 |
| Peña-Mora F. | 5 | 242 | 10 |
| Ryu J. | 5 | 37 | 11 |
| Lee S.H. | 4 | 247 | 8 |
| Li X. | 4 | 34 | 4 |
| Al-Hussein M. | 3 | 4 | 4 |
| Alwasel A. | 3 | 44 | 6 |
| Han S.U. | 3 | 140 | 6 |
| Seo J. | 3 | 94 | 6 |
| Yu Y. | 3 | 105 | 3 |

## Refences